\documentstyle[epsbox,12pt]{article}
\newcommand{\simgt}{\lower.5ex\hbox{$\; \buildrel > \over \sim \;$}}
\newcommand{\simlt}{\lower.5ex\hbox{$\; \buildrel < \over \sim \;$}}
\begin{document}
\title{The Earth effects on the supernova neutrino spectra}
\author{Keitaro Takahashi$^{\rm a}$, Mariko Watanabe$^{\rm a}$, 
Katsuhiko Sato$^{\rm a,b}$\\
{\it $^{\rm a}$Department of Physics, University of Tokyo,
7-3-1 Hongo, Bunkyo,}\\ {\it Tokyo 113-0033, Japan}\\
{\it $^{\rm b}$Research Center for the Early Universe, 
University of Tokyo,}\\
{\it 7-3-1 Hongo, Bunkyo, Tokyo 113-0033, Japan}}
\maketitle

\baselineskip=24pt

\begin{abstract}

The Earth effects on the energy spectra of supernova neutrinos are studied. We analyze
numerically the time-integrated energy spectra of neutrino in a mantle-core-mantle 
step function model of the Earth's matter density profile. We consider a realistic 
frame-work in which there are three active neutrinos whose mass squared differences
and mixings are constrained by the present understanding of solar and atmospheric 
neutrinos. We find that the energy spectra change for some allowed mixing
parameters. Especially, the expected number of events at SNO shows characteristic
behavior with respect to energy, i.e. a great dip and peak.
We show that observations of the Earth effect allow us to identify
the solar neutrino solution and to probe the mixing angle $\theta_{2}$.

\noindent
$PACS:$ 14.60.Pq; 14.60.Lm; 96.40.Tv; 97.60.Bw; 

\noindent
$Keywords:$ Neutrino oscillation; Supernovae; Earth effects;

\end{abstract}
\newpage

\section{Introduction}

Neutrino oscillation is widely accepted as a solution to the solar and atmospheric
neutrino problems. However,
in spite of the remarkable progress made in the studies of solar and atmospheric neutrinos,
there are some ambiguities about neutrino mass spectrum and mixing. Then neutrino emission
from collapse-driven supernova attracts a significant attention because the supernova
neutrinos can probe whole neutrino mass and their studies will help to identify
the pattern of neutrino mass and mixing. \cite{rf:Dighe}
In fact, the direct observations of neutrinos from SN1987A by 
the Kamiokande \cite{rf:Kamioka} and the IMB detectors \cite{rf:IMB} have provided rich
information not only on the mechanism of supernova but also on the neutrino physics 
\cite{rf:Arafune,rf:Minakata} and paved the way for a new phase in the neutrino astrophysics. 
\cite{rf:Sato}

The neutrinos, which are produced in the high dense region of the iron core, \cite{rf:Totani}
interact with 
matter before emerging from the supernova. The presence of non-zero masses and mixing
in vacuum among various neutrino flavors results in strong matter dependent effects,
including conversion from one flavor to another. Hence, the observed neutrino flux
in the detectors may be dramatically different for certain neutrino flavors and for certain
values of mixing parameters, due to neutrino oscillation. \cite{rf:Dighe,rf:Dutta}

Similar matter effects occur in the Earth, too (in connection with solar neutrino
and SN1987A, see \cite{rf:Hata,rf:Fogli} and \cite{rf:Yanagida,rf:Nunokawa,
rf:Jegerlehner,rf:Lunardini}, respectively). The neutrino trajectory inside the 
Earth before reaching the detector depends on the direction of the supernova relative
to the Earth and the time of the day. The comparison of signals from different detectors
would allow one to reveal the Earth matter effects. Also certain features of the energy
spectra can reveal the Earth matter effect even from the observations in one detector.
In this paper, we study numerically the Earth 
matter effects on supernova neutrino spectra with all three active neutrinos
considered. We show that some ambiguities in the solar neutrino solutions
and in $\theta_{2}$ can be resolved by the Earth effect of supernova neutrino.

\section{Three-flavor formulation}

We consider three-flavor mixing among active neutrinos. The three flavor 
eigenstates are related to the three mass eigenstates in vacuum through
a unitary transformation,
\begin{equation}
|\nu_{\alpha}\rangle = \sum_{a=1}^{3} U_{\alpha a} |\nu_{a}\rangle,
\end{equation}
where $|\nu_{\alpha}\rangle$ are flavor eigenstates with flavor $\alpha 
(\alpha = e, \mu, \tau)$ and $|\nu_{a}\rangle$ are mass eigenstates with mass $m_{a}$. 
A convenient parameterization for $U = U(\theta_{1},\theta_{2},\theta_{3})$ is given by
\begin{equation}
U=
\left(\begin{array}{ccc}
c_{2} c_{3}                      & s_{3} c_{2}                      & s_{2}       \\
-s_{3} c_{1} - s_{1} s_{2} c_{3} & c_{1} c_{3} - s_{1} s_{2} s_{3}  & s_{1} c_{2} \\
s_{1} s_{3} - s_{2} c_{1} c_{3}  & -s_{1} c_{3} - s_{2} s_{3} c_{1} & c_{1} c_{2}
\end{array}\right), 
\end{equation}
where $s_{i} \equiv \sin \theta_{i}$ and $c_{i} \equiv \cos \theta_{i}$ for $i=1,2,3$.
The quantities $\theta_{i}$ are the vacuum mixing angles. We have here put the CP
phase equal to zero in the CKM matrix. 

Experimental evidence suggests \cite{rf:SK,rf:SOUDAN,rf:MACRO} that both solar and atmospheric
neutrinos display flavor oscillations, and hence that neutrinos
have mass. All the existing experimental results on the 
atmospheric neutrinos can be well described in terms of the
$\nu_{\mu} \leftrightarrow \nu_{\tau}$ vacuum oscillation with
mass squared difference and the mixing angle given by 
\cite{rf:Fukuda}

\begin{equation}
\Delta m^{2}_{atm} \approx 7 \times 10^{-3} {\rm eV}^2 \; , \;   
\sin^{2} 2 \theta \approx 1 .
\end{equation}  

In contrast, the current (mean event rate) solar neutrino data admit three types
of MSW solutions if the solar $\nu_{e}$ undergo two-neutrino transitions into active
neutrinos, $\nu_{e} \leftrightarrow \nu_{\mu}$ \cite{rf:solar}: the SMA, the LMA and
the so-called ``LOW'' solution. We use SMA and LMA parameters as samples.
\begin{eqnarray}
({\rm SMA}) \;\Delta m^{2}_{\odot} & \approx & (4 \sim 10) \times 10^{-6}  eV^2 \; , \;
 \sin^{2} 2 \theta_{\odot} \approx (2 \sim 10) \times 10^{-3} \\
({\rm LMA}) \;\Delta m^{2}_{\odot} & \approx & (1 \sim 10) \times 10^{-5}  eV^2 \; , \; 
 \sin^{2} 2 \theta_{\odot} \approx 0.7 \sim 0.95 
\end{eqnarray}
These ambiguities mentioned above can be resolved by supernova data including the Earth 
effects. \cite{rf:Dighe}

\section{Calculating the Earth effects}

The calculations of the energy spectra of neutrinos performed in this paper
are based on the time-evolution-operator method developed by Ohlsson and Snellman.
\cite{rf:Ohlsson} This method provides the time evolution operator in terms of the
mass squared differences and the vacuum mixing angle without introducing the
auxiliary matter mixing angles. The evolution operator $U_{f}(L,A,E)$ is a $3 \times 3$
matrix whose components are functions 
of the length $L$ of the neutrino trajectory, the matter density parameter $A$ and
the neutrino energy $E$. The neutrino conversion probability amplitude is given by
\begin{equation}
A_{\nu_{\alpha} \rightarrow \nu_{\beta}} = \langle \nu_{\beta}| U_{f}(L,A,E) |\nu_{\alpha} \rangle ,
\end{equation}
where $A \equiv G_{F} \rho / \sqrt{2} m_{N}$, $\alpha, \beta = e, \mu, \tau$,
$G_{F}$ is the Fermi constant, $m_{N}$ is
the nucleon mass, and $\rho$ is the matter density.

When the neutrinos travel through a series of matter densities with matter density
parameters $A_{1}, A_{2}, \cdots, A_{n}$ and thicknesses $L_{1}, L_{2}, \cdots, L_{n}$,
the total evolution operator is simply given by
\begin{equation}
U_{f}^{tot} = U_{f}(L_{n},A_{n},E) \; \cdots \; U_{f}(L_{2},A_{2},E) \; U_{f}(L_{1},A_{1},E)
\end{equation}
We use the following values as the matter density parameter of the Earth core and mantle, 
respectively;
\begin{eqnarray}
A_{core} = 4.35 \times 10^{-13} {\rm eV} \; , \; (\rho_{core} = 11.5 {\rm g/cm^3}) , \\
A_{mantle} = 1.70 \times 10^{-13} {\rm eV} \; , \; (\rho_{mantle} = 4.5 {\rm g/cm^3}) .
\end{eqnarray}

When neutrinos from such a distance as in the case of supernova arrive at the
Earth, they are not coherent and interference between the different
mass eigenstates does not occur \cite{rf:Dighe}. This effect can be included
in our calculation by time-averaging the conversion probabilities at the
surface of the Earth.

In order to compute the neutrino energy spectra, we used energy spectra of 
neutrinos arriving at the surface of the Earth (Fig.\ref{fig:1}) calculated by 
Watanabe et al. \cite{rf:Mariko}
They have been studying the future detection of a supernova neutrino burst taking
the three-flavor neutrino oscillation inside the star into account. 
The system has two resonances:
one at higher density (H-resonance) and the other at lower resonance (L-resonance).
Adiabaticity at these resonances significantly influences on not only the neutrino flux
outside the star but also the magnitude of the Earth effects.

We computed energy spectra of $\nu_{e}$ and $\bar{\nu}_{e}$ for the
following sets of mixing parameters,
$$\begin{array}{|c|c|c|c|c|c|}\hline
  & \sin^{2} 2 \theta_{1} & \sin^{2} 2 \theta_{2} & \sin^{2} 2 \theta_{3} 
& \Delta m_{12}^{2}({\rm eV}^{2})  & \Delta m_{23}^{2}({\rm eV}^{2}) \\ \hline 
(a) &  0.9  & 0.04             & 0.8              & 2.7 \times 10^{-5} & 7.0 \times 10^{-3}  \\ \hline   
(b) &  0.9  & 0.04             & 7 \times 10^{-3} & 5.0 \times 10^{-6} & 7.0 \times 10^{-3}  \\ \hline  
(c) &  0.9  & 1 \times 10^{-6} & 0.8              & 2.7 \times 10^{-5} & 7.0 \times 10^{-3}  \\ \hline  
(d) &  0.9  & 1 \times 10^{-6} & 7 \times 10^{-3} & 5.0 \times 10^{-6} & 7.0 \times 10^{-3}  \\ \hline  
\end{array}$$
where $m_{1} < m_{2} < m_{3}$ (normal hierarchy) and $\Delta m_{12}^{2}
= m_{2}^{2} - m_{1}^{2} $, $\Delta m_{23}^{2}= m_{3}^{2} - m_{2}^{2} $.
The value of $\theta_{2}$ determines the adiabaticity at the H resonance.
\cite{rf:Dighe} The conversion is adiabatic and nonadiabatic for 
$\sin^{2} 2 \theta_{2} = 0.04$ and $1 \times 10^{-6}$ respectively. 
$\sin^{2} 2 \theta_{3} = 0.8$ and $7 \times 10^{-3}$ correspond to the solutions to the solar 
neutrino problem LMA and SMA, respectively.

In terms of the classification by Dighe and Smirnov \cite{rf:Dighe}, the parameters
(a), (b), (c) and (d) correspond to I-LMA, I-SMA, III-LMA and III-SMA, respectively.
They predict that the Earth effect would be observed in spectra of $\bar{\nu}_{e}$ of (a),
both $\nu_{e}$ and $\bar{\nu}_{e}$ of (c) and $\nu_{e}$ of (d).
In our calculation, on the contrary, significant Earth effects ($> 10 \%$) can be seen in three among $2 
\times 4$ spectra stated above: $\bar{\nu}_{e}$ of (a) and both $\nu_{e}$ and $\bar{\nu}_{e}$ of (c), 
not $\nu_{e}$ of (d).

The flux reaching the detector from a supernova at a distance $d$ from the Earth is 
reduced by an overall geometric factor of $1/(4 \pi d^{2})$. We used $10$ kpc as the
value of $d$.
We computed the flux of neutrinos at 130 discrete energies in a mantle-core-mantle 
step function model of the Earth's matter density profile. For each neutrino
flux with specific energy, we performed a sequence of energy averaging operations
as follows. First we compute the $\nu_{\alpha} \rightarrow \nu_{\beta}$ conversion
probabilities $P_{\nu_{\alpha} \rightarrow \nu_{\beta}}(E)$ at $0.1$ MeV intervals. 
Then the flux at $E$ MeV at the detector $F^{D}(E)$ is,
\begin{equation}
F^{D}_{\nu_{e}}(E) = \sum_{flavor} [F_{\nu_{\alpha}} \sum_{E'=E-1 {\rm MeV}}^{E+1 {\rm MeV}}  
P_{\nu_{\alpha} \rightarrow \nu_{e}}(E')],
\end{equation}
where $F_{\nu_{\alpha}}$ is the $\nu_{\alpha}$ flux just arriving at the Earth.
We take the energy width $2$ MeV taking a rough energy resolutions of detectors 
into account. In fact, the energy resolutions of detectors depend on energy it self.
But the rough estimation of energy resolution is sufficient, since our purpose is 
to see overall shapes of the spectra.

\section{Results and discussions}

\begin{figure}[h]
\begin{center}
\psbox{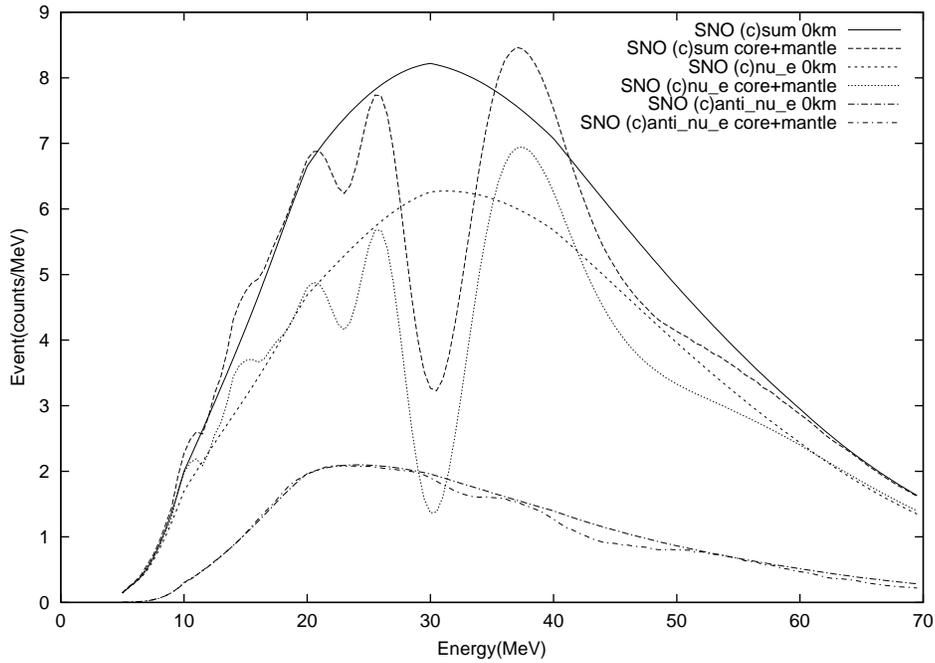}
 \caption{Time-integrated energy spectra at SNO with parameter (c). 
          Short-dashed and dotted line correspond to ${\nu}_{e}$ events 
          without and with the Earth effects, respectively. 
          Long-dashed-dotted and short-dashed-dotted line correspond to 
          ${\bar{\nu}}_{e}$ events without and with the Earth effects, 
          respectively. Solid and long-dashed line correspond 
          to total events without and with the Earth effects, respectively.}
 \label{fig:2}
\end{center}
\end{figure}

We calculated energy spectra of neutrino which traveled through the mantle($3000$ km)-core($7000$ km)
-mantle($3000$ km), which correspond to zero nadir angle. Significant Earth effects
were seen in the flux of ${\nu}_{e}$ with mixing parameter (c) and ${\bar{\nu}}_{e}$
with mixing parameter (a) and (c). Then we converted the fluxes in which 
the Earth effects can be seen to energy spectra at SNO and SuperKamiokande.
In calculating the energy spectra at SNO, two charged current interactions,
\begin{eqnarray}
{\nu}_{e} + d & \rightarrow & p + p + e^{-} \\
{\bar{\nu}}_e + d & \rightarrow & n + n + e^{+}
\end{eqnarray}
are taken into account. For the cross sections for these reactions, we refer to \cite{rf:Ying}.
The detection efficiency is taken to be one.
Fig.\ref{fig:2} shows the time-integrated energy spectra of 
${\nu}_{e}$,${\bar{\nu}}_{e}$ and ${\nu}_{e} + {\bar{\nu}}_{e}$ at SNO 
with mixing parameters (c) with and without the Earth effects.

\begin{figure}[h]
\begin{center}
\psbox{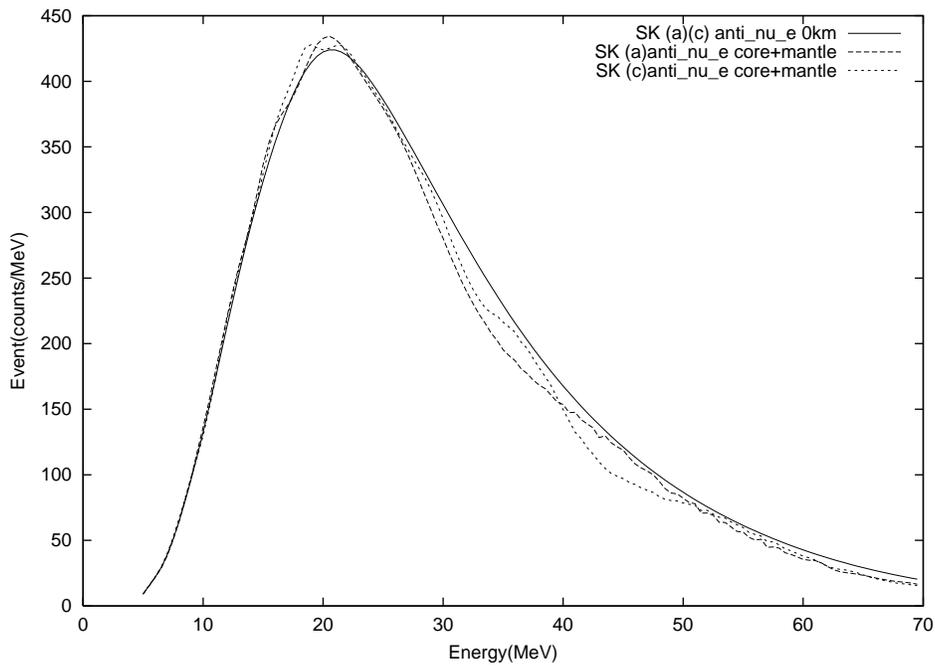}
 \caption{Time-integrated energy spectra at SK. Solid line corresponds 
          to energy spectra without the Earth effect. Dashed and dotted line 
          correspond to energy spectra with Earth effect using parameter (a) 
          and (c), respectively. }
 \label{fig:3}
\end{center}
\end{figure}

In calculating the energy spectra at SuperKamiokande, the reaction,
\begin{equation}
{\bar{\nu}}_e + p \rightarrow n + e^{+},
\end{equation}
which dominates the total event number, is taken into account.
For the cross section of this reaction, we refer to \cite{rf:Totsuka}.
The appropriate detection efficiency curve is also taken into account \cite{rf:efficiency}.
Fig.\ref{fig:3} shows the time-integrated energy spectra at SuperKamiokande 
with mixing parameters (a) and (c) with and without the Earth effects.
We don't show the energy spectra at SNO with mixing parameter (a) 
because the Earth effects on ${\nu}_{e}$ and ${\bar{\nu}}_{e}$ are negligible 
and small, respectively, and the number of event coming from  ${\nu}_{e}$ is dominant.
For mixing parameter (b) and (d), no significant Earth effect can be seen.

\begin{figure}[h]
\begin{center}
\psbox{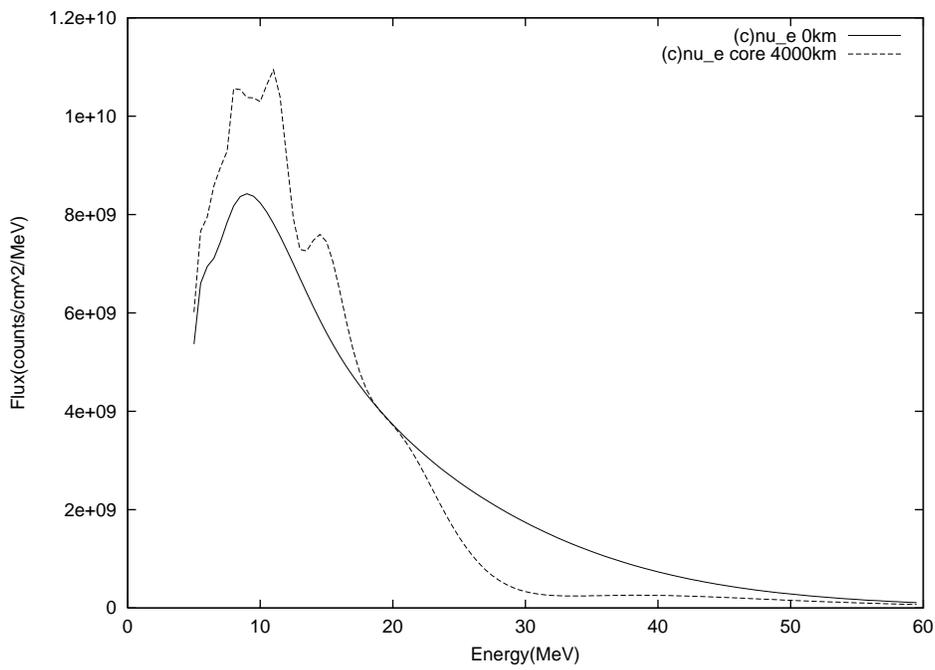}
 \caption{Energy spectrum of $\nu_{e}$ with parameters (c) which did not travel (solid)
          and traveled through the core 4000km (dashed). }
 \label{fig:4}
\end{center}
\end{figure}

\begin{figure}[h]
\begin{center}
\psbox{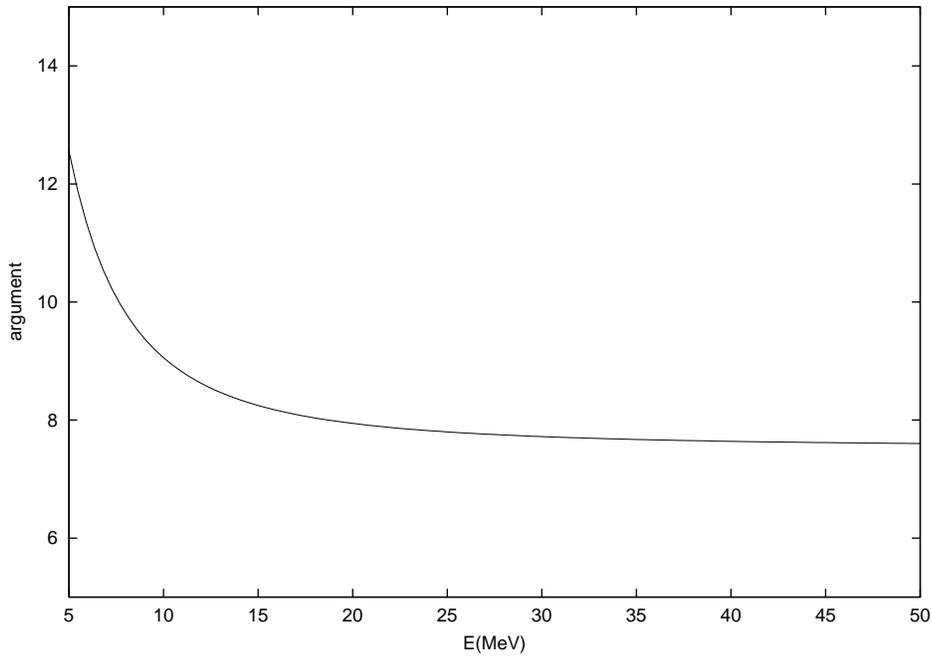}
 \caption{Argument of the second factor of the probability $P_{\nu_{e} \rightarrow \nu_{\mu}}$ }
 \label{fig:5}
\end{center}
\end{figure}

The flux of the neutrino passing through the Earth oscillates with respect to the energy. 
This behavior can be explained as follows.
For simplicity, we assume the density of the Earth be homogeneous. The dashed line in Fig.\ref{fig:4}
shows the flux of $\nu_{e}$ which traveled $4000$ km through the Earth core. Probability of
neutrino conversion, for example, $\nu_{e} \rightarrow \nu_{\mu}$ can be expressed 
as a function of the length the neutrino trajectory in matter and the neutrino energy,
\cite{rf:Bahcall}
\begin{equation}
P_{\nu_{e} \rightarrow \nu_{\mu}} = \sin^{2} 2 \theta_{M} \sin^2 \frac{\pi R}{L_{M}(E)},
\end{equation}
where $L_{M}(E)$ is the matter oscillation length which can be written with the
vacuum oscillation length $L_{V}$, vacuum mixing angle $\theta_{V}$, and the neutrino-electron
interaction length $L_{e}$ as
\begin{equation}
L_{M}(E) = \frac{L_{V}}{(1-2 \frac{L_{V}}{L_{e}} \cos 2 \theta_{V} + (\frac{L_{V}}{L_{e}})^{2})^{1/2}},
\end{equation}
where $L_{V} = 4 \pi E \hbar / \Delta m^{2} c^{3}$, 
$L_{e} = \sqrt{2} \pi \hbar c / G_{F} n_{e}$, and $n_{e}$ is the electron number density. Substituting,
\begin{equation}
R = 4000 {\rm km} \; , \; \cos 2 \theta_{V} = 0 \; , \; n_{e} = n_{core},
\end{equation}
we obtain the argument of the second factor of $P_{\nu_{e} \rightarrow \nu_{\mu}}$,
$\frac{16 \sqrt{1+0.0225 E^{2}}}{E} \pi$. This function decreases from $13$ to $8$ as
$E$ increases from $5$ MeV to $15$ MeV and takes an almost constant value $8$ for $E > 15$ MeV:
the oscillation length asymptotically approaches a constant value. (Fig. \ref{fig:5}) 

\begin{figure}[h]
\begin{center}
\psbox{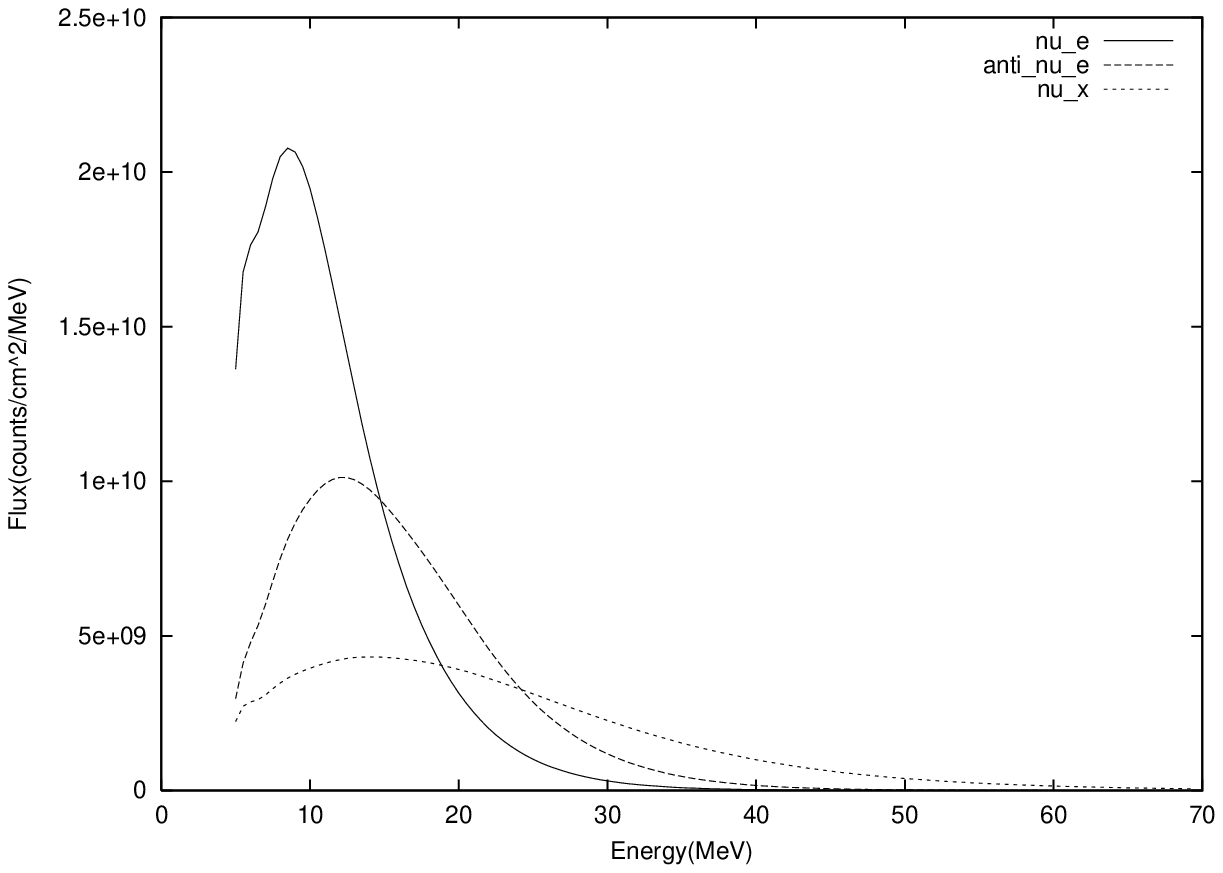}
 \caption{Energy spectra of $\nu_{e}$, $\bar{\nu}_{e}$ and $\nu_{x}$ of the 
          numerical supernova model used in this paper. The solid, dashed and 
          dotted lines are the spectrum of $\nu_{e}$, $\bar{\nu_{e}}$ and 
          $\nu_{x}$, respectively.}
 \label{fig:1}
\end{center}
\end{figure}

On the other hand, the difference between the $\nu_{e}$ flux $F^{D}_{e}$ at the detector and
the $\nu_{e}$ flux $F_{e}$ before traveling the Earth is proportional to, \cite{rf:Dighe}
\begin{equation}
F^{D}_{e} - F_{e} \propto F^{0}_{e} - F^{0}_{x},
\end{equation}
where $F^{0}_{e}$ and $F^{0}_{x}$ are the original fluxes of, respectively, $\nu_{e}$ and
$\nu_{\mu,\tau}$ at the center of the supernova, which vanish at $E \approx 20$ MeV. \cite{rf:Mariko}
Fig.\ref{fig:1} shows the original fluxes: the solid, dashed and dotted lines show $F^{0}_{e}$, 
$F^{0}_{\bar{e}} $and $F^{0}_{x}$, respectively.

The shapes of spectra can be understood roughly by the latter factor. At lower energy 
$F^{0}_{e} - F^{0}_{x} > 0$, then $F^{D}_{e} - F_{e} > 0$. $F^{0}_{e} - F^{0}_{x} = 0$
at $E \approx 20$ eV , then $F^{D}_{e} - F_{e} = 0$. At higher energy $F^{0}_{e} - F^{0}_{x} < 0$,
then $F^{D}_{e} - F_{e} < 0$. The small oscillation can be understood as the oscillation of 
$P_{\nu_{e} \rightarrow \nu_{\mu}}$.

\begin{figure}[h]
\begin{center}
\psbox{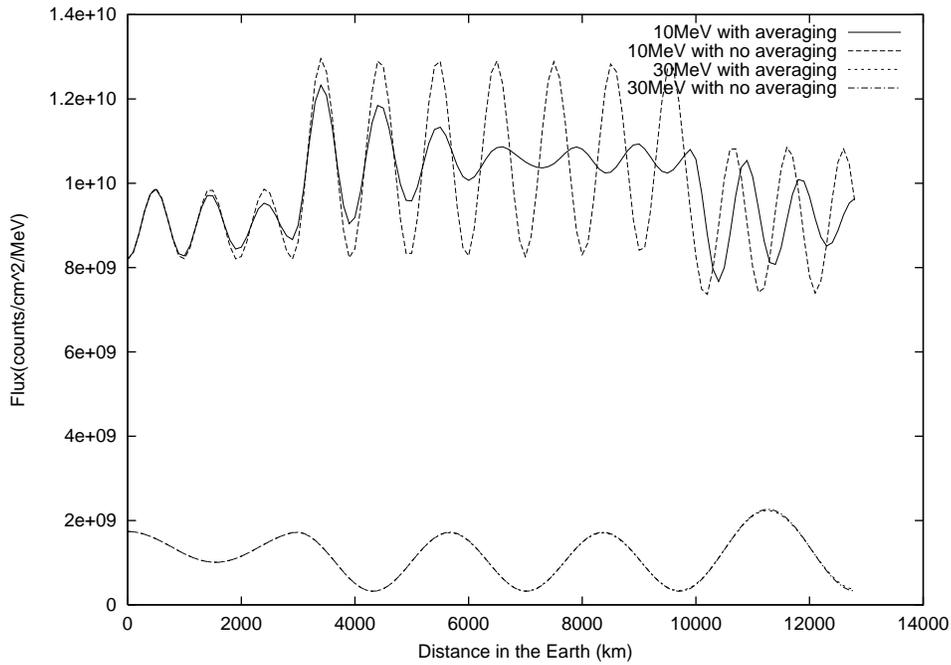}
 \caption{$\nu_{e}$ flux in the Earth. The solid and dashed lines show $\nu_{e}$ flux at $10$ MeV without and
          with the averaging operations, respectively. The dotted and dash-dot-dash lines show $\nu_{e}$ flux
          at $30$ MeV without and with the averaging operations, respectively (almost degenerate). }
 \label{fig:6}
\end{center}
\end{figure}

Fig.\ref{fig:6} shows the $\nu_{e}$ flux inside the Earth with parameters (c) which
travel the mantle, core, and mantle. The length of the mantle and core are set to be
$3000$ km and $7000$ km, respectively.
If averaging operations are performed, flux oscillations damp because
$\nu$ flux of different energies have different oscillation lengths
and cancel the oscillation behavior while traveling long distance.
Since the oscillation lengths are comparable to the radius of the Earth,
the Earth effects are strongly dependent on the neutrino path inside the Earth.
So the magnitude of the Earth effects will be different at each detector. 
      
Significant Earth effects can be seen in the energy spectra at SNO for 
mixing parameter (c) (Fig.\ref{fig:2}). There are a large dip at 30 MeV 
and a peak at 37 MeV.
Binning appropriately, (for example $25 \sim 35$ MeV and $35 \sim 40$ MeV, etc.)
we can tell whether the Earth effects exist. Since the energy value of dip and peak
change if zenith angle of neutrino path changes,
appropriate binning will change accordingly.

The Earth effects in Fig.\ref{fig:3} are very small.
This smallness comes partly from the smallness of $F^{0}_{\bar{e}} - F^{0}_{x}$ (Fig.\ref{fig:1}).
Although the expected number of events at SuperKamiokande is very large 
($\sim 10000$ counts)\cite{rf:Mariko} and then statistical error would be small,
it would be difficult to detect the Earth effect on neutrinos from supernova at 10kpc
from the Earth due to the uncertainties in supernova model. 
In this case we suggest as a criterion of the Earth effect the ratio of events at 
lower energy ($< 20$ MeV) to that at higher energy ($> 20$ MeV).
These subjects need quantitative and statistical argument, which we plan to
study in great detail elsewhere.\cite{rf:Takahashi}

If we observe a supernova and detect neutrinos in future, whether the Earth effects exist 
can help to determine the neutrino parameters. 
\begin{itemize}
\item When the Earth effects are observed at both SuperKamiokande and SNO,
the neutrino parameter is (c).
\item When the Earth effects are observed at SuperKamiokande only,
the neutrino parameter is (a).
\end{itemize}

Recent result of solar neutrino analysis at SuperKamiokande favor LMA rather 
than SMA. \cite{rf:Suzuki} If the solution of the solar neutrino problem would turn out to be
LMA, the Earth effects on supernova neutrino can distinguish parameter (a)
from (c): we could determine ${\theta}_{2}$, which is only loosely constrained
by the other current experiment.

\section{Acknowledgments}

K.Takahashi would like to thank G.Watanabe and S.Nagataki for useful discussions. 
This work was supported in part by
Grants-in-Aid for Scientific Research provided by the Ministry of Education,
Science and Culture of Japan through Research Grant No.07CE2002.

\newpage

\end{document}